\documentclass[aps,pre,reprint,groupedaddress]{revtex4-1}
\usepackage{graphicx}
\usepackage{dcolumn}
\usepackage{color}
\usepackage{hyperref} 
\usepackage{bm}

\begin{document}


\title{Signal integration enhances the dynamic range in neuronal systems}


\author{Leonardo L. Gollo}
\email[]{leonardo@ifisc.uib-csic.es}
\author{Claudio Mirasso}
\author{V\'{\i}ctor M. Egu\'{\i}luz}

\affiliation{IFISC (CSIC - UIB), Instituto de F\'{\i}sica
  Interdisciplinar y Sistemas Complejos, E-07122 Palma de Mallorca, Spain}


\date{\today}

\begin{abstract}
The dynamic range measures the capacity of a system to discriminate the intensity of an external stimulus. 
Such an ability is fundamental for living beings to survive: to leverage resources and to avoid danger. 
Consequently, the larger is the dynamic range, the greater is the probability of survival. 
We investigate how the integration of different input signals
affects the dynamic range, and in general the collective behavior of a network of excitable units. 
By means of numerical simulations and a mean-field approach, 
we explore the nonequilibrium phase transition in the presence of integration. 
We show that the firing rate in random and scale-free networks undergoes a discontinuous phase transition depending on both the integration time and the density of integrator units. Moreover, in the presence of external stimuli, we find that a system of excitable integrator units operating in a bistable regime largely enhances its dynamic range. \end{abstract}

\pacs{}

\maketitle



\section{Introduction} A system operating in the vicinity of a critical
state can present several advantages. For instance, hair cells of the auditory system poise themselves close to a Hopf bifurcation~\citep{Eguiluz00,*Camalet00}, and in neuronal systems
it has been proposed to provide optimal solutions for sensory stimuli
detection~\citep{Kinouchi06,Shew09}, the transmission and storage of
information~\citep{Beggs03, Beggs04, *Beggs08, *Plenz07, *Haldeman05,
  *Hsu06}, and computational capabilities~\citep{Legenstein07}. These
results motivated discussions of how the brain can, if it does,
operate in a critical state and whether it could be due to
self-organization arguments~\citep{Levina07} or by evolutionary
reasons~\citep{Bonachela10}. Neural systems operating in a critical
state also provide an alternative explanation of how the brain
integrates the activity of distant regions~\citep{Beggs03}. In the
critical regime, the correlation length diverges and neurons from
different areas can effectively share information. Based on these arguments
and on experimental evidences~\citep{Fraiman09}, it has been suggested
that the brain should be tuned around a critical point of a
second-order phase transition to efficiently process
information~\citep{Fraiman09, Chialvo04,*Chialvo08,*Kitzbichler09,*Werner10, *Chialvo10,*Tagliazucchi11}.

Excitable media have been proved to serve as excellent stimulus
intensity processors. Their fundamental nonlinear interactions of
excitable waves confer a great capacity to compress several decades
of stimulus intensity inputs into a single decade of firing rate
output~\citep{Gollo12}. This capability, which has also
been proposed to be the main function of neuronal active
dendrites~\citep{Gollo09}, is robust
for different networks~\citep{Furtado06, *Wu07,*Copelli07,
 *Assis08,Kinouchi06, Gollo09, Larremore11}.
In many contexts, such as gene regulatory networks~\citep{Szejka08}, and neuronal~\citep{Kandel} and social
systems~\citep{Dodds04,*Granovetter78,*Watts02,*Centola07,*Centola10}, the typical elementary unit dynamics results
from the integration of neighbor contributions. In neuroscience, it
remains a fundamental open problem to understand how a singular
membrane potential output is generated by the convergence of complex
spatio-temporal synaptic integration~\citep{Magee00,
  Gollo09,Gollo12}. To accrue for this difficulty, neurons present
a myriad of active channels~\citep{Reyes01}, dendritic
structures (even within the same neuron type~\citep{Spruston08}), and
temporal integration modes. For example, the efficacy of the
presynaptic neurons is largely variable, and neurons might require up
to hundreds of excitatory postsynaptic potentials to
spike~\citep{Kandel}.

In this letter we demonstrate that integration of excitable units is a central element to shape the dynamics of the system:
The nonequilibrium phase transition, between the resting and the self-sustained configurations,
switches from a continuous second-order to a discontinuous first-order transition.
Along with this discontinuity, a history-dependent bistable phase emerges.
In this phase, the input-output response changes and the dynamic range is strikingly enhanced.
We show the generality of the result with respect to the network topology, the integration time window, and the number of input signals needed to fire.
Moreover, we point out how the presence of a bistable phase changes the paradigm of maximum dynamic range at criticality~\citep{Kinouchi06}.
Such an optimum regime typically appears in the bistable regime and depends on the past history.

\begin{figure*}[t]
\includegraphics[width=6.75in]{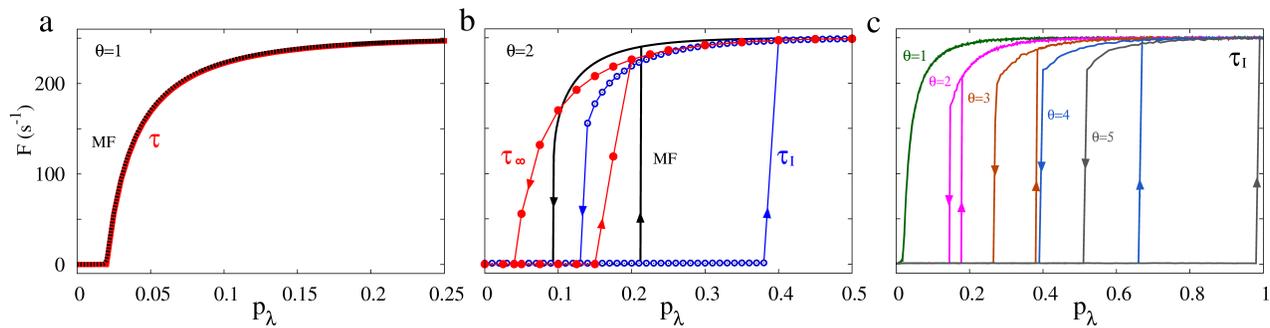}
\caption{\label{fig1} Continuous and discontinuous spontaneous
  activity $F$ versus $p_\lambda$.
  Mean-field approximation (MF) and numerical results for
  random networks of (a) non-integrators ($\theta=1$), 
  (b) integrators ($\theta=2$) for both $\tau_I$ and $\tau_\infty$ integration times
  with $N=5,000$; and (c) integrators with different threshold values, for $\tau_I$ and $N=1,000$. 
 Other parameter values are $\Delta p_\lambda=0.0025$ and $F_0=3\%$. }
\end{figure*}

\section{The model} As a simple and influential excitable media,
we explore the Kinouchi-Copelli model~\citep{Kinouchi06, Greenberg78} generalized to account for the integration of multiple excitatory inputs.
We consider $N$ nodes embedded in sparse (Erd\H{o}s-R\'{e}nyi) random
and (Barab\'{a}si-Albert) scale-free networks~\citep{Albert02},
both with an average degree $K=50$.
Each node $i$ represents an excitable unit whose state $s_i(t)\in \{0,1,2\}$ indicates whether the unit is in the quiescent state [$s_i(t)=0$],
in the active state [$s_i(t)=1$], or in the refractory state [$s_i(t)=2$].
The dynamics obeys probabilistic rules with a synchronous update, and $\delta t\equiv 1$~ms is the discrete time step.
Every node $i$ at time $t$ updates its state as follows:
\begin{itemize}
\item In the active state $s_i(t)=1$, it switches to the refractory state $s_i(t+\delta t)=2$;
\item In the refractory state $s_i(t)=2$, it returns to the quiescent state $s_i(t+\delta t)=0$ with probability $p_\gamma=\frac{1}{2}$;
\item Nodes in the quiescent state $s_i(t)=0$ become active either (i) by an external driving
(or spontaneous activation) with probability $p_h=1-\exp(-h \delta t)$ per time step,
where $h$ is the rate of a Poisson process; or (ii) by the integration of the contributions received from their active neighbors, with probability $p_\lambda$.
\end{itemize}
In order to model the integration process,
we count the number of neighbor contributions $\Lambda_i(t)$ received within the time window of width
$\tau$: $(t-\tau, t)$.
In the absence of external driving,
a node $i$ spikes if $\Lambda_i$ reaches at least $\theta$ inputs,
\textit{i.e.}, $\Lambda_i(t) \geq \theta$.
Two extreme limits of integration time are of particular interest: the
{\em infinite integration time} $\tau \rightarrow \infty$ ($\tau_\infty$), where the
integration window takes into account the entire current quiescent
history of the node, and a {\em coincidence detection} $\tau =  1$ ms, ($\tau_I$), where the integration time is limited to $\delta t$.

\section{Continuous versus discontinuous phase transition} In the absence of external driving ($h=0$),
the standard model without integration ($\theta=1$) leads to a continuous phase transition~\citep{Kinouchi06}.
The average firing rate $F$,
calculated over all nodes and over a large time window ($10$ s),
grows smoothly for increasing coupling strength above the critical value $p_{\lambda}^c$ (Fig.~\ref{fig1}a).
The critical point is determined by the largest eigenvalue of the network adjacency matrix~\citep{Larremore11}.
For a random network (Fig.~\ref{fig1}a), the critical value is $p_{\lambda}^c=K^{-1}$, when the average number of spikes induced by each spike (branching ratio) is one~\citep{Kinouchi06}.
Conversely, in the presence of integration ($\theta >1$) the phase transition occurs abruptly,
generating a bistable phase with a hysteresis cycle (see the mean-field approach below).
We calculated the hysteresis cycles by varying $p_\lambda$ upward and downward
along the whole range in small steps of $\Delta p_\lambda$, activating at each change of $p_\lambda$
a small fraction of nodes ($F_0$, from $1\%$ to $3\%$) to allow the system
to escape from the resting configuration.
As shown in Fig.~\ref{fig1}b, the change in the nature of the phase transition is observed for any value of the integration time,
as well as in the mean-field approximation.
The discontinuous phase transition is also robust for any value of $\theta>1$,
illustrated in Fig.~\ref{fig1}c for $\tau_I$.
It can be also seen from the figure that larger threshold values generate larger hysteresis cycles.

\begin{figure}[t]
\includegraphics[width=3.1in]{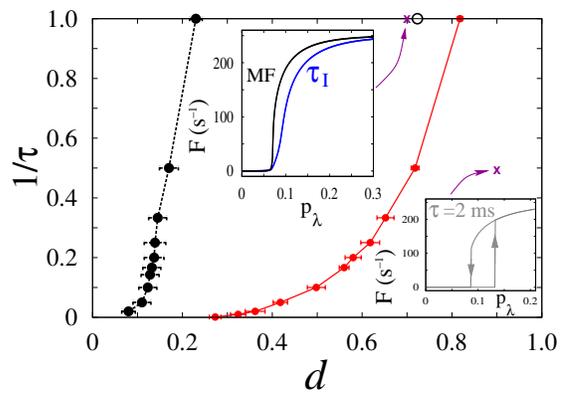}
\caption{\label{fig2} Dependence of the nature of the phase transition order on the
  integration time $\tau$ for networks ($N=5,000$) composed of a
  mixture of both integrators (with a density $d$ of $\theta=2$ nodes)
  and non-integrators. The solid (dashed) line corresponds to the random (scale-free) network
  for $\Delta p_\lambda=0.001$ and $F_0=1\%$.
  The left-hand side of the curve corresponds to a 
  continuous phase transition whereas the right-hand side corresponds to a 
  discontinuous phase transition.  The error bars correspond to the standard
  deviation over ten trials.
  The black open symbol depicts the mean-field
  shift in the order of the phase transition.  The left inset panel compares
  the mean-field approximation with the simulations for the density of
  integrators $d=70\%$ and $\tau_I$.  The right inset panel illustrates a
  discontinuous phase transition for $\tau =2$ ms and $d=90\%$.  }
\end{figure}

\begin{figure*}[t]
\includegraphics[width=7.15in]{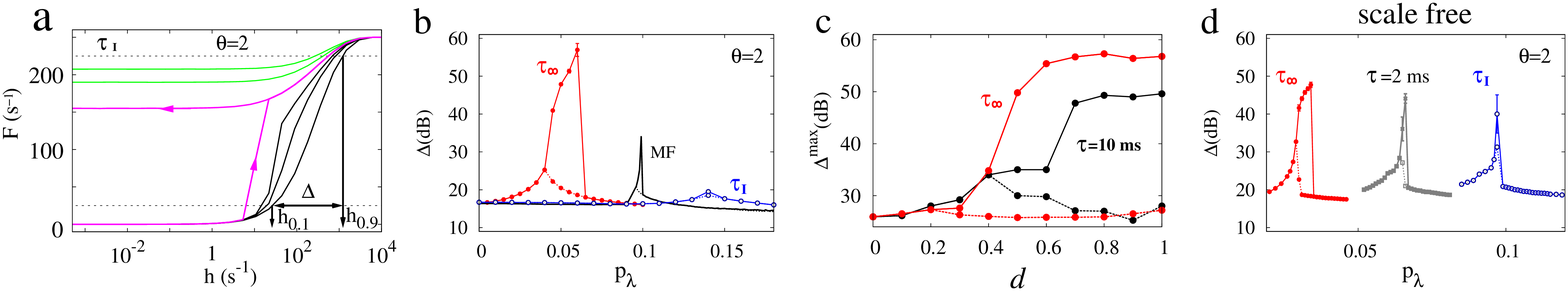}
\caption{\label{fig3} Response curves and dynamic range in networks
  of integrators ($\theta=2$) with $N=5,000$.  (a) Family of response
  functions for a random network with $\tau_I$ and (from right to left) $p_\lambda=0, 0.1, 0.12,
  0.14, 0.16, 0.18$. (b) and (d) Dynamic range versus coupling strength for
  (b) random and (d) scale-free networks.
  At the bistable region, the dotted line (bottom) stands for initial conditions with a 
  high activity level and the continuous line (top) stands for initial conditions
  with a low activity level. The error bars correspond to the standard
  deviation over six trials of two realizations. (c) Dependence of the maximum dynamic range $\Delta^{\rm max}$ 
  on the density of integrators in random networks.}
\end{figure*}


While the previous analysis assumes identical nodes, next 
we consider heterogeneous populations composed of both integrators ($\theta=2$)
and non-integrators ($\theta=1$) nodes.
This situation corresponds to the intermediate configuration between integrators, as in
Fig.~\ref{fig1}b, and non-integrators, as in Fig.~\ref{fig1}a.
For random and scale-free networks,
the minimum density of integrator nodes ($d$)
that yields a discontinuous phase transition depends on the integration time scale $\tau$,
as shown in Fig.~\ref{fig2}. Although in both cases the density of integrators needed to display a discontinuous phase transition decreases with increasing integration time, the scale-free network
requires a lower density of integrators.
The integration time is fundamental to bind the collective dynamics together.
Coincidence detection restricts the scope of action of the integrator nodes and
the network is effectively split in two parts according to the threshold values.
For example, in a random network
with $\tau_I$ and a density of integrators below $80\%$,
the dynamics is dominated by the sub-group of active non-integrators,
leading to a continuous phase transition.
In this case of continuous transition,
the integrator nodes do not interfere much in the dynamics:
The effective connectivity is $K (1-d)$,
and the expected critical point for the phase transition is given by
$p_{\lambda}^c \simeq \frac{1}{K (1-d)}$
(for the left inset panel of Fig. 2: $K =50$, and $d=0.7$, $p_{\lambda}^c=\frac{1}{15}$).
For larger integration times ($\tau > \tau_I$),
the discontinuous phase transition (as exemplified by the right inset panel) gradually dominates,
and the right side of the transition increases with $\tau$.
In this case, the integrator nodes, although spiking less, tend to remain active,
furnishing clear influence in the collective dynamics.
Therefore, the prevailing dynamics carries the integrators finger-print
given by the discontinuous phase transition.

\section{Dynamic range} So far we have analyzed the behavior of the excitable media  in the absence of external stimuli.
In the remainder, we are interested in the response of the system as a function of the external driving,
considered as a Poisson process with rate $h$.
As illustrated in Fig.~\ref{fig3}a,
the response functions for different coupling $p_\lambda$ grow with external driving rate $h$ and saturate at
a maximum firing rate $F_{\rm max}=\frac{1}{2+p_\gamma^{-1}}$ ms$^{-1}$, which is 
determined by the refractory period $p_\gamma$.
Among the response functions, there are three regimes. For very low
coupling, the response functions are subcritical, the self-sustained
solution is not allowed, and the activity dies out when $h\rightarrow
0$. On the high coupling limit, small perturbations lead the system to
the self-sustained mode. In between both regimes there is the bistable
region. Response functions in this regime are history dependent. Very
small perturbations are typically not enough to drive the system to
the self-sustained mode. However, at a certain external driving rate
(which is trial dependent) the system becomes active, as depicted by
the upward arrow in Fig.~\ref{fig3}a. On the contrary, if the
response function is calculated by reducing the external driving
(leftward arrow), the system maintains a high firing rate and the
activity does not die out when $h\rightarrow 0$. This path dependence
could explain the large fluctuations in the experimental response
functions as well as the dependence on the measurement time
period in the olfactory system~\citep{Friedrich97, *Wachowiak01, *Bhandawat07}.

The bistable regime also confers path dependence to the dynamic range. 
Figure~\ref{fig3}a depicts the key elements of the standard dynamic
range definition. The two horizontal dashed lines stand for $F_{0.1}$
(bottom) and $F_{0.9}$ (top). They correspond to $10\%$ and $90\%$
of the maximum firing rate ($F_{\rm max}$) subtracted from the minimum
firing rate [$F(h\rightarrow 0)$], and they cross the response
functions respectively at the external driving intensities of $h_{0.1}$ and $h_{0.9}$.
The dynamic range is thus defined as the number of decades comprised between $h_{0.1}$ and $h_{0.9}$:
$\Delta \equiv10 \log \frac{h_{0.9}}{h_{0.1}}$.  Figures~\ref{fig3}b and~\ref{fig3}d
show the dynamic range for networks of integrators with different
integration times $\tau$, for random and scale-free networks. 
In the bistable regime,
when a high firing rate is observed (bottom line), the system is only able to distinguish the input level intensity. 
For a low firing rate (top line), the system not only distinguishes the input intensity
but also detects the abrupt change in the firing rate. 
The system displays the largest dynamic range in the low firing rate and the maximum appears in the bistable regime. 
The height and width of the peak of the dynamic range curves depend on the integration time. 
Coincidence detectors show a poor capacity to distinguish the incoming input 
(lower peak and narrower width of $\Delta$ as a function of $p_\lambda$). 
However, for large enough density of integrators, the dynamic range increases with longer integration times 
(see Fig.~\ref{fig3}c), which increases the capacity to discriminate incoming inputs.
Table~\ref{tableb} compares the dynamic range of the neuronal networks with and without integrators: 
The maximum enhancement of the dynamic range as a consequence of the collective behavior 
[\textit{i.e.}, $\Delta^{\rm max}-\Delta(p_\lambda=0)$] is over four times larger in the presence 
(than in the absence) of integration in both random and scale-free networks.
\begin{table}[ht!]
\begin{center}
\begin{tabular}{|c|c|c|c|c|} \hline
&\multicolumn{2}{|c|} {non-integrators ($\theta=1$)} & \multicolumn{2}{|c|} {integrators ($\theta=2$, $\tau_\infty$)}  \\ \cline{2-5}
\vspace{-0.3 cm}
 &   &  &  &   \\
network & $\Delta^{\rm max}-\Delta(0) $ & $\Delta^{\rm max}$ & $\Delta^{\rm max}-\Delta(0)$ & $\Delta^{\rm max}$ \\ \hline \hline
random &  10  & 26 & 41 & 57 \\
scale-free &  7  & 23 & 32 & 48 \\
\hline
\hline
\end{tabular}
\end{center}
\vspace{-0.5 cm}
\caption {Dynamic range (dB) for network size $N=5,000$.}
\label{tableb}
\end{table}

\section{Mean-field approach} 
\begin{figure}[!]
\includegraphics[width=3.3 in]{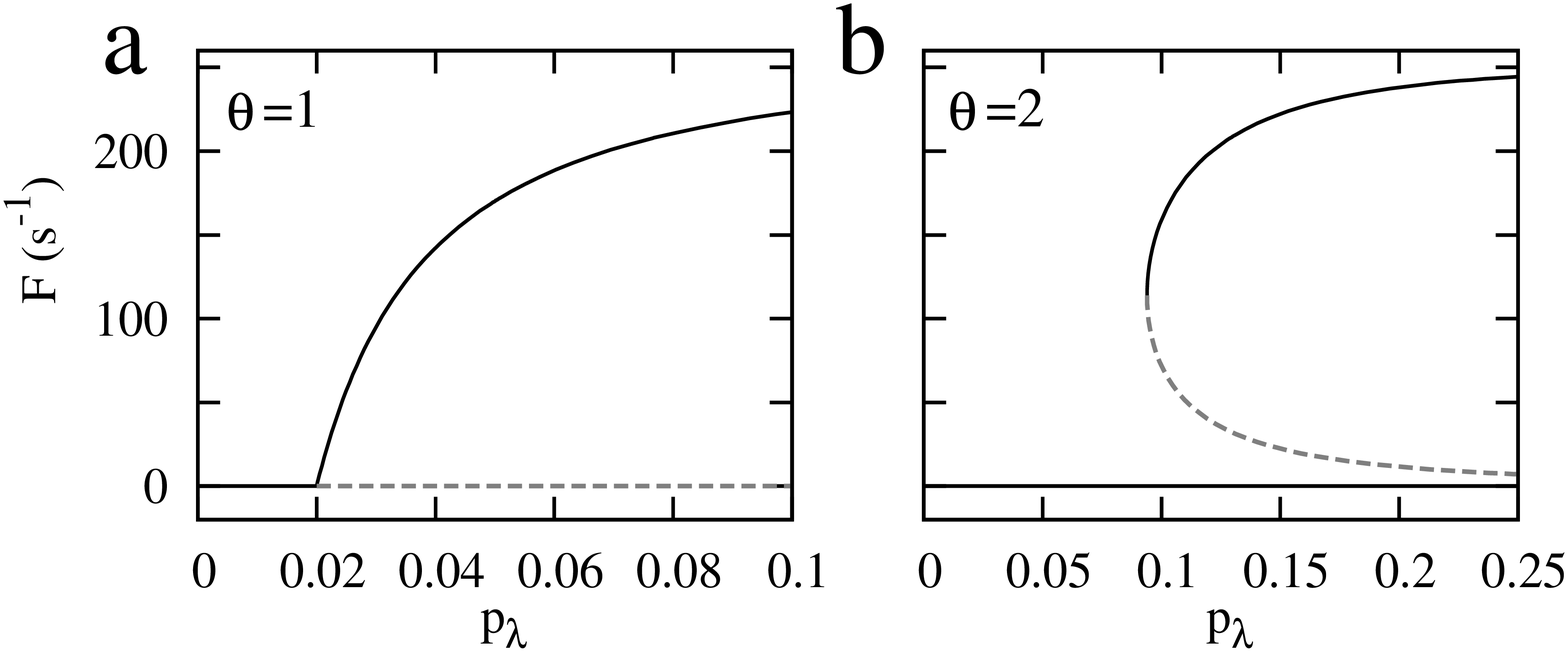}
\caption{\label{fig4} Bifurcation diagram of the mean-field approximation: (a) transcritical for $\theta=1$; (b)
saddle node for $\theta=2$. Solid lines, stable stationary solutions; dashed lines, unstable ones.}
\end{figure}

The mean-field approximation we present corresponds to the
model version for coincidence detection ($\tau_I$). For $K
\gg \theta$, the mean-field map for the average firing rate $F$ of a
population with threshold $\theta$ can be written as:
\begin{equation}
\label{homogeneous}
\delta_t F_{t+1}= Q_t p_h +  Q_t
(1-p_h) \Lambda_t^{\theta}  \, ,
\label{eq1}
\end{equation}
where $\Lambda_t^{\theta}=[ 1-(1-p_\lambda \delta_t F_t)^{K} ]^{\theta}$
is the probability of a quiescent node to become active in the next
time step due to at least $\theta$ neighbor contributions within a
single time step, $Q_t=1-\delta_t F_t-R_t$ is the probability of finding a site in the quiescent state, and
$R_{t+1}= \delta_t F_t+(1-p_\gamma) R_t$ is the probability of finding a site in the refractory state.
Iterating the map until convergence we get the solution of $F$ in the
stationary configuration ($t\rightarrow \infty$), which is used to compare
with the simulations.  The numerical solutions of Eq.~(\ref{eq1}) for various conditions are shown in Figs.~\ref{fig1} and~\ref{fig3}. For the population of non-integrators
(Fig.~\ref{fig1}a) we recover the Kinouchi-Copelli
equation~\citep{Kinouchi06}, which describes particularly well the
behavior in random networks.  In the presence of integration, the
result captures qualitatively the behavior of the phase transitions
(Fig.~\ref{fig1}b), the response function, and the dynamic range (Fig.~\ref{fig3}b). 
A bifurcation analysis reveals some aspects of the
phase transition as a function of the threshold $\theta$.
In the absence of input ($h=0$), as shown in Fig.~\ref{fig4}, 
for $\theta=1$ there is a transcritical bifurcation;
for  $\theta>1$ a saddle-node bifurcation and a stable fixed point at $F=0$ coexist.

Analogously, one can also extend the results for heterogeneous populations, as
considered in Fig.~\ref{fig2}. 
At any time~$t$, we define, for each subpopulation of threshold $\theta_i$,  
$F_t^{(\theta_i)}$, $R_t^{(\theta_i)}$, and $Q_t^{(\theta_i)}$
as the firing rate, and the probability of finding a site in the refractory and in the quiescent states, 
respectively. The firing rate of the network is
given by $F_t=F_t^{(1)}  (1-d) +  F_t^{(2)} d$, where $d$ denotes the density
of integrator nodes. Then, by generalizing Eq.~\ref{homogeneous} we can find $F^{1}$ and $F^{2}$ from:
\begin{eqnarray}
\delta_t F_{t+1}^{(\theta_i)} &=& Q_t^{(\theta_i)} p_h +  Q_t^{(\theta_i)} (1-p_h) \Lambda_t^{\theta_i}  \, .
\label{eq2}
\end{eqnarray}
As shown in the left inset panel of Fig.~\ref{fig2}, the average firing rate of the network 
qualitatively captures the phase transition.

\section{Summary and conclusions} We have studied the collective behavior of an excitable media where the units integrate incoming signals~\citep{Kandel}.
The presence of a minimum density of integrator nodes leads the system to an abrupt phase transition. 
Discontinuous transitions have been observed experimentally and in threshold models~\citep{Breskin06,*Cohen10}, in models with adaptive interactions~\citep{Levina09,Millman10},
and in the presence of strong nonlinear coupling~\citep{Assis09}.

As a consequence of the discontinuous phase transition, bistability emerges. 
In the context of neuroscience, bistability is known to play an important role in memory maintenance~\citep{Fuster71}. 
A bistable regime composed of a configuration with high or low activity levels~\citep{Steriade93} has also been observed in cortical neurons. 
Since most neurons (if not all) must integrate their incoming post-synaptic potentials,
our results suggest that the transition to the regime of self-sustained activity
in a neuronal system could be restricted to a discontinuous transition type.

Concerning the output response to external stimulus (which might vary for orders of magnitude),
the bistable regime provides two different response types, depending on the history
(either with low or high activity levels for $h\sim 0$). 
The low past activity level with an infinite integration time gives rise to the largest
dynamic range in random and scale-free networks. 
Taking this finding into account,
biologically inspired artificial stimulus detectors
with great capabilities can be designed from excitable media composed of integrator units \citep{Medeiros11}. 
Moreover, we expect that our results might also be relevant to other systems where integration plays an important role as, for instance, in gene regulatory networks~\citep{Szejka08}, and
social interaction~\citep{Granovetter78,*Watts02,*Centola07,*Centola10},
and it would be interesting to explore the behavior of the dynamic range in the recently found explosive percolation~\citep{Achlioptas09}.

We thank to Ernest Motbri\'{o}, Miguel A. Mu\~{n}oz, Ernesto M. Nicola, John Rinzel and
Olaf Sporns for useful discussions. LLG acknowledges Prof. John Rinzel and
his working group for the hospitality and valuable discussions during
his visit in 2009 to the Center for Neural Science at New York
University. The authors acknowledge financial
support by grants from the MICINN (Spain) and FEDER under project
FIS2007-60327 (FISICOS) and FIS2011-24785 (MODASS).

\bibliography{./Integrator}

\end{document}